\documentclass{emulateapj}
\usepackage{apjfonts}

\shorttitle{The radio-optical connection in 3C~120}
\shortauthors{Le\'on-Tavares et al.}

\begin{document}
\title{Relativistic plasma as the dominant source of the optical
  continuum emission in the broad-line radio~galaxy
  \objectname{3C~120}} \author{J. Le\'on-Tavares\altaffilmark{1,2, 3$\star$},
  A. P. Lobanov\altaffilmark{1}, V. H. Chavushyan\altaffilmark{2},
  T. G. Arshakian\altaffilmark{1},
  V. T. Doroshenko\altaffilmark{4,5,6}, S. G. Sergeev
  \altaffilmark{5,6}, Y. S. Efimov \altaffilmark{5,6}, and
  S.V. Nazarov\altaffilmark{5} }

\altaffiltext{1}{Max-Planck-Institut f\"ur Radioastronomie,
                 Auf dem H\"ugel 69, 53121 Bonn, Germany}

\altaffiltext{2}{Instituto Nacional de Astrof\'{\i}sica \'Optica y
   Electr\'onica, Apartado Postal 51 y 216, 72000 Puebla, M\'exico
}

\altaffiltext{3}{Aalto University Mets\"ahovi Radio Observatory,
  Mets\"ahovintie 114, FIN-02540, Kylm\"al\"a, Finland}

\altaffiltext{4}{Crimean Laboratory of the Sternberg Astronomical
  Institute, P/O Nauchny, Crimea 98409, Ukraine}

\altaffiltext{5}{Crimean Astrophysical Observatory, P/O Nauchny,
Crimea 98409, Ukraine}

\altaffiltext{6}{Isaac Newton Institute of Chile, Crimean Branch}

\altaffiltext{$\star$ }{leon@kurp.hut.fi}

\begin{abstract}
  We report a relation between radio emission in the inner jet of the
  Seyfert galaxy \objectname{3C~120} and optical continuum emission in
  this galaxy.  Combining the optical variability data with
  multi-epoch high-resolution very long baseline interferometry
  observations reveals that an optical flare rises when a superluminal
  component emerges into the jet and its maxima is related to the
  passage of such component through the location a stationary feature
  at a distance of $\approx$1.3\,parsecs from the jet origin. This
  indicates that a significant fraction of the optical continuum
  produced in \objectname{3C~120} is non-thermal and it can ionize
  material in a sub-relativistic wind or outflow. We discuss
  implications of this finding for the ionization and structure of the
  broad emission line region, as well as for the use of broad emission
  lines for determining black hole masses in radio-loud AGN.
\end{abstract}

\section{INTRODUCTION}

In the current astrophysical paradigm for active galactic nuclei
(AGN), each constituent of an AGN contributes to a specific domain in
the spectral energy distribution (SED).  The SED of some AGN, show a
significant amount of energy excess at UV/optical wavelengths, which
is commonly attributed to be produced in an accretion disk around the
putative central supermassive black hole (BH). Surrounding the
accretion disk at $\lesssim 1$ pc, there is a central broad line
region (BLR) formed by high density gaseous clouds orbiting the
central BH. The thermal UV/optical emission radiated from the disk is
thought to be the prime source of variable optical continuum and a
dominant factor for the ionization of the BLR material. In radio-loud
AGN however, the broad-band continuum can also have a substantial
contribution from non-thermal synchrotron radiation generated in the
relativistic jet (D'Arcangelo et al. 2007, Marscher et al. 2008, Soldi
et al. 2008). To understand the mechanism and properties of the
broad-line generation in such objects, it is pivotal to be able to
localize and identify the region where the bulk of the non-thermal
optical continuum emission is produced.

An efficient way to identify a region responsible for the production
of the variable optical continuum emission in radio-loud AGN is to
combine long-term optical monitoring and regular very long baseline
interferometry (VLBI) observations. A link between the optical
emission and relativistic jet in 3C~120 was suggested by Belokon
(1987), based on a correlation found between the optical outburst and
ejections of plasma clouds (jet components) in the jet. Using the
radio and optical data available from 14 years monitoring of the radio
galaxy 3C\,390.3, Arshakian et al. (2008, 2009) found that ejections
of new components can be associated with optical flares occurring on
time scales from months to years and reaching their maxima during
passages of the jet components through a stationary emitting region in
the inner jet, at a distance of $\sim 0.5$\,pc from the jet
origin. This indicates that the variable optical continuum emission
can be generated in the innermost part of the jet, in the region
located upstream from the stationary feature. This has an important
implication for the existence of a non-virialized outflowing
broad-line region associated with the jet.

\begin{figure*}[t!]
   \includegraphics[width=\textwidth]{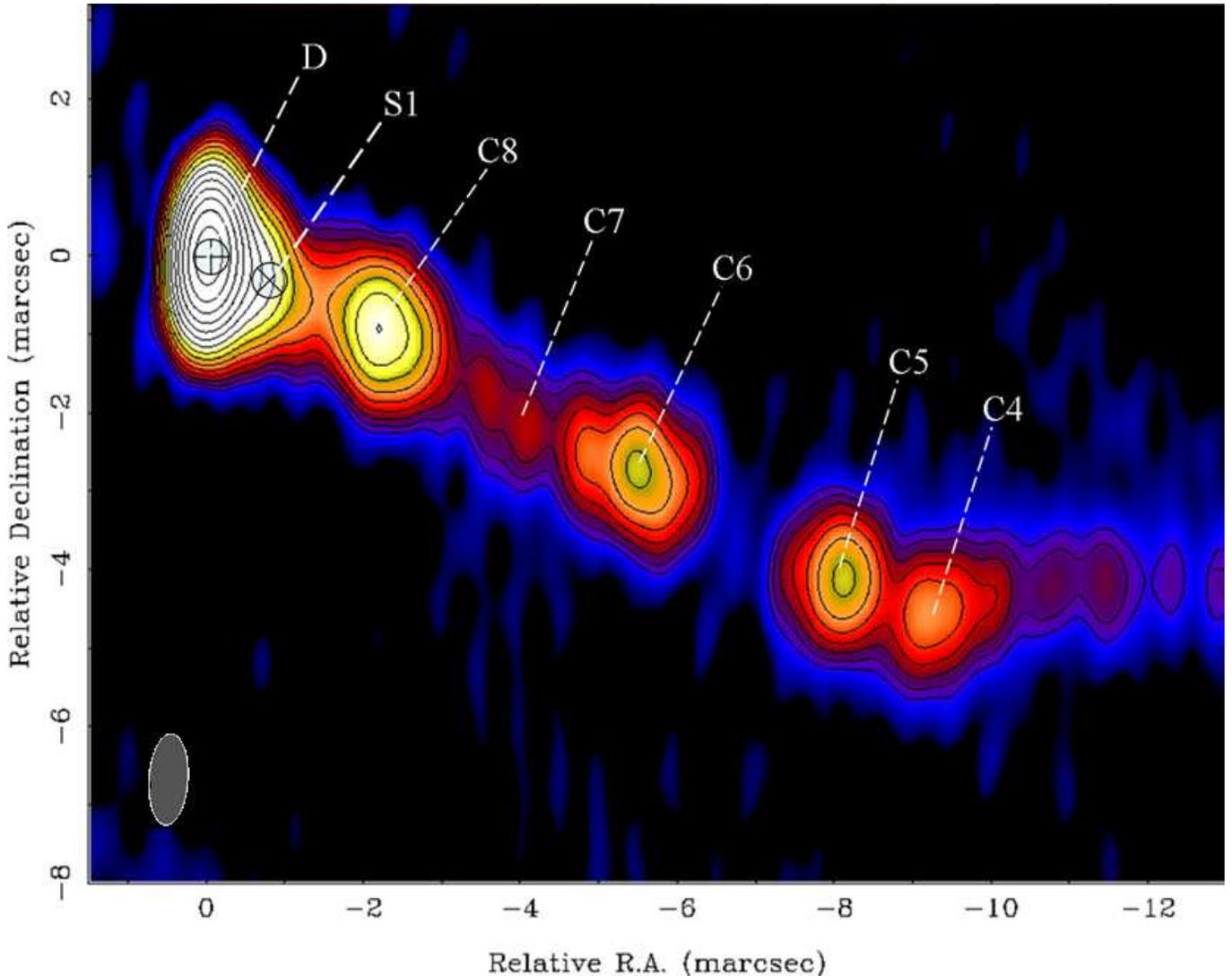}
\caption{ Compact jet in 3C~120 observed on January 7, 2005, with the
  VLBA at 15~GHz (2cm). Shaded ellipse in the lower left corner marks
  the restoring point-spread function (beam) with the FWHM of 0.53
  $\times$ 1.26 mas oriented at an angle of three degrees (clockwise
  rotation). The peak flux density in the image is 974 mJy/beam and
  the rms noise is 0.26 mJy/beam. Contours are drawn at 1, $\sqrt{2}$,
  2...\% of the lowest contour shown at 0.9
  mJy/beam. The jet structure is parametrized by a set of
    two-dimensional, circular Gaussian features obtained from fitting
    the visibility amplitudes and phases (Pearson 1997). The model-fit
    components (C4-C12) located within the inner 14 mas from the core
    are shown in Figure \ref{Fig:3c120_kinematic_model}.  Label marks
    show the location of the model-fit components and the two
    stationary components D and S1 are highlighted by open circles.}
\label{Fig:3c120_jet}
\end{figure*}

It should be noted that stationary and low-pattern speed features have
been recently reported in parsec-scale jets of a number of prominent
radio sources ({\em e.g.}, Kellermann et al. 2004, Savolainen et
al. 2006, Lister et al. 2009b), alongside faster, superluminally
moving components embedded in the same flows (Kellermann et al. 2004,
Jorstad et al. 2005). Although specific geometric conditions and
extremely small viewing angles may lead to formation of such features
in relativistic flows (Alberdi et al. 2000), it is more likely that
these features represent standing shocks (for instance, recollimation
shocks in an initially over-pressurized outflow; G\'omez et al. 1995,
Perucho \& Mart\'i 2007). Such standing shocks may play a major role
in accelerating particles near the base of the jet (Mandal \&
Chakrabarti 2008; Becker et al. 2008), and could be responsible for
the persistent high levels of polarization in blazars (D'Arcangelo et
al. 2007; Marscher et al. 2008).

It is important to understand whether the correlations found in
Arshakian et al. (2008, 2009) are characteristic only for the
3C\,390.3 or common for all radio galaxies and quasars. In this
letter, we combined archival VLBI monitoring data and data available
from photometric monitoring of 3C\,120 to study the link between
properties of the superluminal jet and optical continuum flares.

Throughout the paper, a flat $\Lambda$CDM cosmology is assumed, with
the Hubble constant H$_{0}= 70$ km s$^{-1}$ Mpc$^{-1}$ and matter density
$\Omega_{m}=0.3$. This corresponds to a linear scale of 0.658\,pc/mas,
at the redshift $z=0.033$ of \objectname{3C~120}.

\section{OBSERVATIONS}

To match the available optical monitoring data, twenty
VLBA\footnote{Very Long Baseline Array of the National Radio Astronomy
  Observatory, USA} observations of \objectname{3C~120} made between
2001 December 23 and 2008 November 26 have been retrieved from the
MOJAVE (Monitoring of Jets in Active Galactic Nuclei)
database. Details of the observations and reduction of the MOJAVE data
are presented in Kellermann et al. (1998,2004); Lister \& Homan
(2005); Lister et al.  (2009a).  Individual images of the VLBA
experiments analyzed in this work can be found at the MOJAVE
website\footnote{http://www.physics.purdue.edu/MOJAVE/sourcepages/0430+052.shtml}.
VLBI images for all twenty epochs have been produced, applying
standard self-calibration and hybrid imaging procedures (cf., Cornwell
\& Fomalont 1999) to uniformly weighted and untapered visibility data.

The optical CCD-BVRI observations used in this paper were carried out
at the 70-cm telescope of the Crimean Astrophysical Observatory from
2002 January 05 to 2008 March 11. The details about instrument, device
and observations are described in Sergeev et al. (2005).  The CCD
camera Ap7p was mounted on the prime focus (f=282 cm) of the
telescope. The CCD is provided for B,V,R,R1,I filters, where R1 filter
closely resembles the Cousins I filter, while the other filters
matches closely to the standard Johnson filters. We used the light
curve in B filter, constructed from aperture photometry through a 15"
diameter aperture centered on the galaxy nucleus relative to a
comparison star No. 1 in the field (Doroshenko et al. 2005).

\section{STRUCTURE AND KINEMATICS OF THE JET}

The 15~GHz radio structure of \objectname{3C~120} shown in Figure
\ref{Fig:3c120_jet} is characterized by a one-sided jet directed at a
P.A. of about $-120^{\circ}$. Two dimensional circular Gaussian
components were fitted in the (\textit{u,v})-domain to the fully
calibrated visibility data using the modelfitting technique (cf.,
Pearson 1997).  

Robustness of the modelfitting is verified by comparing the noise
properties and the peak and total flux densities corresponding to the
Gaussian component models to those from the respective hybrid
images. This is done by using the procedure described in the appendix
of Arshakian et al. (2009). We find that, in all our model fits, these
parameters are similar to those from the respective hybrid images,
which indicates that the Gaussian model fits represent adequately the
structure of the source.

Self-consistency of the component identification is verified by
requiring a smooth kinematic and flux density evolution of each of the
individual components, similarly to the methods employed in other
works dealing with long-term VLBI monitoring data (cf., Lobanov 1996,
Lobanov \& Zensus 1999). We restrict our analysis to a full kinematic
model for the inner 14 milliarcseconds (mas) of the jet, since beyond
that distance a reliable model could not be established due to the
lower surface brightness and high complexity of the outer region. We
identify nine moving features (C4-C12) moving down the jet at
superluminal speeds. Estimates of errors of the integrated flux
density, size and position of each component in the inner jet were
calculated using the method given by Fomalont~(1999), taking into
account the resolution limits of the observations (Lobanov
2005).

\begin{figure}[t!]
\begin{center}
\resizebox{\hsize}{!}{
   \includegraphics[trim = 0cm 0cm 0cm 0cm,angle=0]{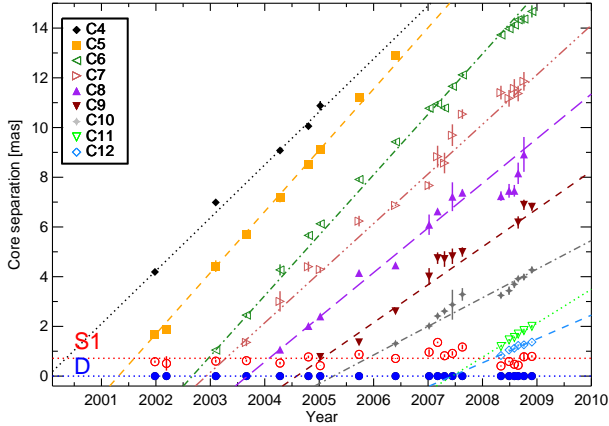}
}
\end{center}
\caption{\label{Fig:3c120_kinematic_model} Relative separations of the
  jet components from the stationary feature D (blue circles) for 20
  epochs of the VLBA observations of 3C~120. Moving components are
  denoted by their respective label marks. A stationary component S1
  is denoted by red open circles. The lines represent the best linear
  least-squares fits to the component separations. Errors
    smaller than symbols may not be visible.}
\end{figure}

In all twenty epochs, we identify a stationary component (S1)
separated from D by $0.715 \pm 0.253$ mas, corresponding to the
projected linear separation of $0.471 \pm 0.133$ pc.  Adopting an
angle of 20.5$^{\circ}$ (Jorstad et al. 2005) between the jet axis and
the line of sight, the corresponding de-projected distance between D
and S1 is $\sim 1.35 \pm 0.38$\,pc.

Fidelity of the position and flux density measurements for stationary
component S1 is verified by comparing them with the resolution limits
calculated from the SNR of the detections (see Lobanov 2005, Arshakian
et al 2010 for details of the procedure). We find that the component
S1 is always resolved and the flux densities of the component D and S1
estimated from the modelfits are uncorrelated, implying that D and S1
are two independent features in the jet.

The apparent positional variations of S1 around its average location
can be explained by a combination of variable optical depth of the
inner part of the jet (represented by the core component D) and by
blending between D and S1 and between S1 and moving jet features. The
magnitude of the positional change of the optically thick core D can
be estimated from the core shift of $\approx 1$\,mas measured between
2.3\,GHz and 8.4\,GHz (Kovalev et al. 2008) and assuming that the
opacity is caused by synchrotron self-absorption (cf. K\"onigl 1981,
Lobanov 1998). In this case, the location of the core D would vary
$\propto S_\mathrm{D}^{2/3}$ and for the flux variability measured in
our data at 15\,GHz this would result in positional variations with an
r.m.s of $\approx 0.2$\,mas. Since the location of the component S1 is
always measured with respect to the component D, this positional
variability is attributed to variations of the position of S1. The
magnitude of positional uncertainty due to blending can be estimated
from the component modelfits following Lobanov (2005) and Arshakian et
al. (2010). For the case of blending between D and S1, we estimate an
uncertainty of about $0.06$\,mas. Thus, the nuclear opacity and D-S1
blending alone account for a positional uncertainty of $\approx 0.21$
mas, which is close to the measured one. The magnitude of positional
uncertainty due to blending between S1 and moving features is
difficult to estimate, but we expect that it should be similar or even
even larger that the uncertainty due to the D-S1 blending. We
therefore conclude that S1 is indeed likely to be a stationary
formation inside the relativistic flow in 3C\,120.

Time evolution of angular separations of moving jet components from
the stationary core (D) is shown in Figure
\ref{Fig:3c120_kinematic_model}.  We fitted the component separation
by linear regressions and used back-extrapolation of the linear fits
to the component separations to calculate apparent speeds, epochs of
ejection from the component D ($t_\mathrm{D}$) and epochs of passage
through the stationary feature S1 ($t_\mathrm{S1}$), for each moving
component. The resulting apparent speeds (in units of the sped of
light, $c$), ejection times and times of passages through the
stationary component S1 are presented in columns (2)-(4) of Table
\ref{Table:3C120_times}. The linear fits to the observed separations
of the components C4-C12 from the component D yield proper motions in
the range of 0.9-2.4 mas/yr, which correspond to apparent speeds of
2.0-5.3\,$c$.

\begin{deluxetable}{lccrrrr}
\tablecolumns{7}
\tabletypesize{\scriptsize}
\tablewidth{0pt}
\tablecaption{\label{Table:3C120_times} 3C~120 jet components properties and parameters of associated optical flares}
\tablehead{
\colhead{Component} & \colhead{t$_\mathrm{D}$} &  \colhead{t$_\mathrm{S1}$}     & \colhead{$\beta_\mathrm{app}$} & \colhead{$t_\mathrm{fl}$} & \colhead{$N_\mathrm{rel}$}
& \colhead{$\tau_\mathrm{fl}$} \\
\colhead{} & \colhead{[yr]}          & \colhead{[yr]}        &  \colhead{[$c$]}            & \colhead{[yr]} & \colhead{} & \colhead{[yr]}                    \\
\colhead{(1)}       & \colhead{(2)}     &  \colhead{(3)}         & \colhead{(4)}                  & \colhead{(5)} & \colhead{(6)} & \colhead{(7)} 
}
\startdata
   C4  &  2000.02  & 2000.35   &   4.56  $\pm$   0.06 & ... & ... & ... \\
   C5  &  2001.31  & 2001.60   &   5.26  $\pm$   0.02 & $<$2001.9  & $<$10   & $>$0.1  \\
   C6  &  2002.67  & 2002.96   &   5.19  $\pm$   0.01 & 2002.99  & 3.10  & 0.18  \\
   C7  &  2002.94  & 2003.30   &   4.28  $\pm$   0.07 & 2003.21  & 1.80  & 0.11  \\
   C8  &  2003.68  & 2004.08   &   3.83  $\pm$   0.03 & 2004.02  & 1.88  & 0.10  \\
   C9  &  2004.58  & 2005.05   &   3.22  $\pm$   0.06 & 2005.02  & 2.52  & 0.07  \\
  C10  &  2005.26  & 2005.88   &   2.46  $\pm$   0.03 & 2005.85  & 3.24  & 0.14  \\
  C11  &  2007.51  & 2008.00   &   3.12  $\pm$   0.10 & 2007.84  & 4.34  & 0.22  \\
  C12  &  2007.43  & 2008.18   &   2.03  $\pm$   0.08 & 2008.10  & 6.66  & 0.16  \\
\enddata
\tablecomments{
Column.~(1): Component identifier;
Column.~(2): $t_\mathrm{D}$ -  time of ejection from  the component D;
Column.~(3): $t_\mathrm{S1}$ -  time of passage through the stationary component S1;
Column.~(4): $\beta_\mathrm{app}$ radial speed in units of the speed of light;
Column.~(5): $t_\mathrm{fl}$ - fitted peak epoch of optical flares associated with
individual jet components;
Column.~(6): $N_\mathrm{rel}$ - relative particle density increase during a flare;
Column.~(7): $\tau_\mathrm{fl}$ - characteristic time scale of a flare.
}
\end{deluxetable}

\section{RELATION BETWEEN THE JET AND THE OPTICAL CONTINUUM}

\begin{figure*}[t!]
\begin{center}
\resizebox{\hsize}{!}{
   \includegraphics[trim = 0cm 0cm 0cm 0cm]{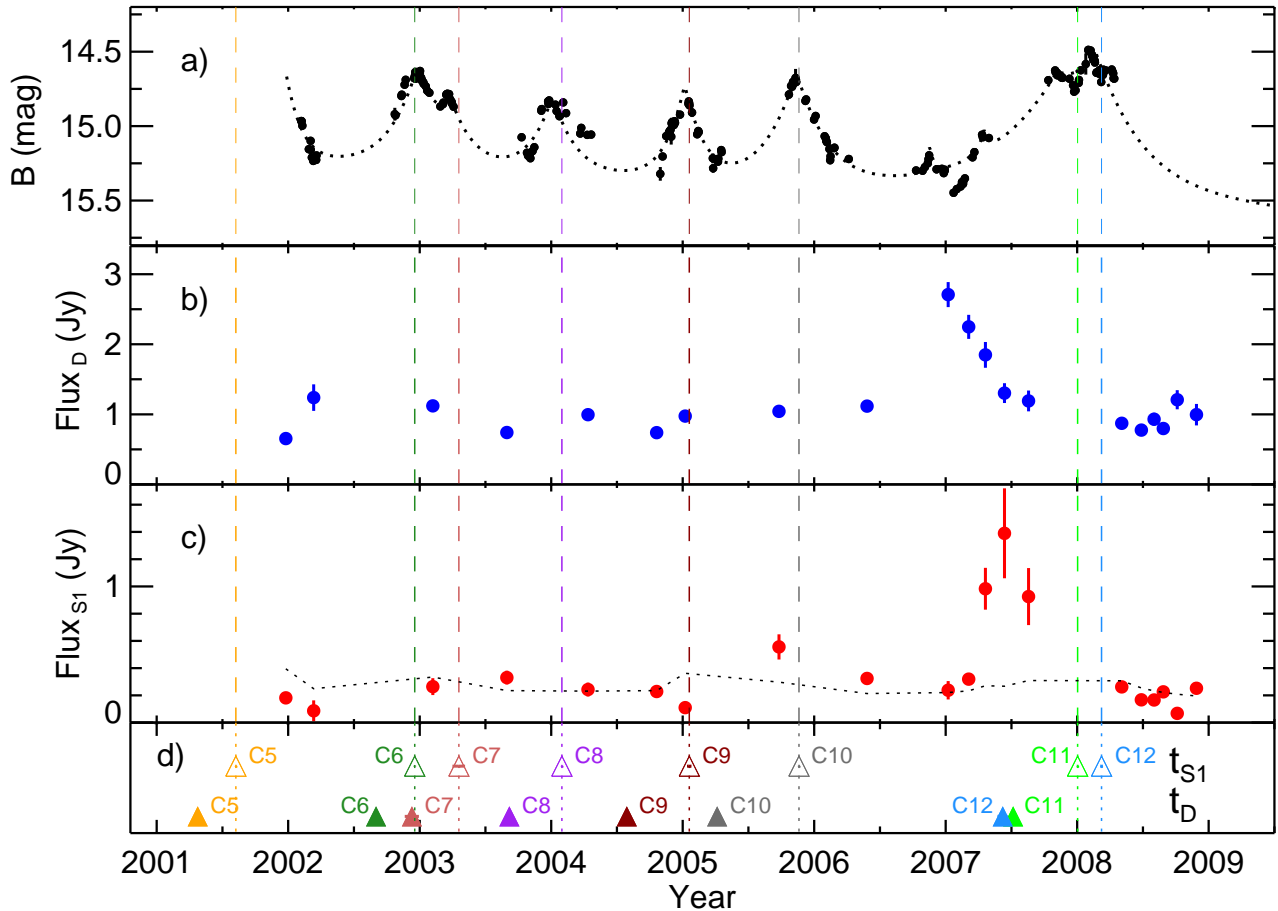}
}
\end{center}
\caption{\label{Fig:3c120_flux} Variability of the optical and radio
emission and jet kinematics in 3C~120 from 2002 to 2008. Panel
description: a)~optical B magnitudes (circles) and flare model (dotted
line) for the optical variations; b) and c)~radio flux densities of
the stationary features D and S1 at 15 GHz (dotted line shows flux
density changes predicted from the model for the optical flares);
d)~times of ejection of radio components at D (filled triangles) and
times of passages of the moving jet components through S1 (open
triangles). The mean time required for the components C4-C12 to move
the distance from D to S1 is $0.45 \pm 0.14$ yr.  Each of the seven
flares in the optical light curve is correlated with a passage of one
of the moving jet features C6--C12 through the location of S1, with an
average time delay less than $ 0.1$ yr between the optical and radio
events.}
\end{figure*}

Relation between the radio and optical emission in \objectname{3C~120}
is presented in Figure~\ref{Fig:3c120_flux}. In this plot, variations
of the optical B-band continuum and radio flux densities of the
stationary features D and S1 at 15 GHz are compared to the ejection
epochs and epochs of passages of the moving components through the
region S1. The optical light curve shows several flares superimposed
on a long-term trend. The sparser radio measurements sample clearly
only one strong flare event around 2007, which is visible in the light
curves of both D and S1.  The epochs of each superluminal ejection
($t_\mathrm{D}$) and passing through the stationary component S1
($t_\mathrm{S1}$) are marked in Figure~\ref{Fig:3c120_flux} by filled
and open triangles, respectively.

Following an ejection from the jet origin at the region D, it takes a
moving jet component $0.45 \pm 0.14$ years to travel the distance
between D and S1. Passages of moving jet components C6-C12 through the
region S1 are correlated with local peaks in the optical light
curve. The radio passages of new components through S1 lag the maxima
of optical flares by an average delay $< 0.1$~yr.  Application of a
statistical test described in Marscher et al. (2002) yields a
probability of $\sim 10^{-9}$ for all 7 radio and optical continuum
events to be so close in the time by chance. However, there is the
possibility that jet components C6-C7 and C11-C12 belong to the same
event, respectively. Thus, the probability that 5 passing times would
be observed to occur randomly within 0.2 yr of the maxima in the
optical flares is $\sim 10^{-5}$. Hence, the correspondence between
$t_\mathrm{S1}$ and maxima of optical flares is still real at a high
confidence level.  This suggests that the region responsible of
optical flares and the stationary component S1 must be physically
connected, similarly to the situation in the radio galaxy
\objectname{3C~390.3} (Arshakian et al. 2009).

We model the optical light curve as synchrotron flares caused by
density variations in the relativistic flows, using the approach
described in Lobanov \& Zensus (1999). We set a quiescent particle
density $N_0 =1$ (as only relative changes of the density are required
for the modelling) and estimate a quiescent magnetic field strength
$B_0 \approx 0.5$\,G from broad band spectral fits (Tyul'bashev \&
Chernikov 2004). We model the optical variations by density variations
with characteristic exponential rise and decay times
$\tau_\mathrm{fl}$. The resulting light curve is shown in
Fig.~\ref{Fig:3c120_flux} and the flare parameters are presented in
Table~\ref{Table:3C120_times} for the events coinciding with the jet
component passages through the region S1. All of the optical flares
require a factor of $\le 7$ increase of the particle density, which is
physically plausible. For a spectral index $\alpha =-0.5$ of the
synchrotron emission, the optical flare fit also yields good
predictions for the radio flux density of S1, with an average ratio of
1.2 between the predicted and measured flux densities, given the fact
that the radio data have not been used at all for producing the
fit. This supports further the suggestions that the optical flares are
related to synchrotron emission originated in the vicinity of the
stationary feature S1.  In 2006.5-2007.5, the optical light curve
shows two minor events peaking close to the peaks of radio emission in
D and S1. Our optical fit indicates that the particle density increase
during these events is very small, offering a speculation that these
flares may result from changes in the magnetic field or Lorentz factor
of the jet plasma.

The strong flare detected in the radio in 2007 is likely related to
the ejection of two moving components, C11 and C12, following an
unusually long period of relative quiescence.  The components C6 and
C7 may also be related to a single event in the nucleus. Paired
ejections of superluminal features have been seen previously in 3C120
(G\'omez et al. 2001) and attributed to trailing shocks in the jet.
The flux density of D component increased almost by a factor of three
during this flare. The flux density of S1 component reacted to the
flare in the component D with a time delay of $\sim 0.4$ years, which
is similar to the travel time of the jet features moving from D to
S1. However, the components C11 and C12 were ejected at the time when
the flare in component D was already fading and the flux density in
component S1 reaches the maximum. This peculiarity may reflect
specific physical conditions in the jet. It may suggest, for instance
that the jet is initially dominated by Poynting flux, and plasma in
the moving features is accelerated by conversion of the Poynting flux
into particle energy, which occurs downstream from the Poynting flux
dominated region. Unfortunately, the present VLBI data are too sparse
to uncover such a degree of detail about the physics of the region
between the features D and S1, and it is not possible to verify
whether the radio flux density of S1 responded to the passage of
moving features in the same fashion as the optical continuum. A
detailed radio-optical study of a single event in 3C~120 (or a similar
object) would be extremely useful for uncovering the complete physical
picture of flaring activity.

\section{DISCUSSION}

The relation between the optical variability and the structural
changes of radio emission in \objectname{3C~120} indicate that a
reliable physical identification of the stationary features D and S1
in the compact jet with respect to the central black hole of
\objectname{3C~120} is instrumental for understanding the radiation
mechanisms and the structure of the region where the bulk of variable
continuum emission is produced.  The correlations presented above
suggest the component D is located at the base of the relativistic
jet, while S1 is most likely a recollimation shock or internal oblique
shock formed in the continuous relativistic flow (cf., Daly \&
Marscher 1988, G\'omez et al. 1995, Romanova \& Lovelace 1996,
Marscher et al. 2002, 2008). Alternatively, the component D could be
the core of the jet located at some distance from the accretion disk
while the component S1 can be identified with a standing
shock. Another possibility is that the components D and S1 may be
constituting the bases of the counterjet and the jet respectively.  In
this case, the component S1, located on the jet side, should be
brighter than the component D because of the relativistic Doppler
effect. It is evident from Figure~\ref{Fig:3c120_flux} however that D
was the brighter feature over the entire period of the VLBI
monitoring. This counter-jet nature of D is further rejected by the
later rise of the flux density of S1 during the flare in 2007 -- while
S1 should have flared before the component D, if the latter was
located on the counter-jet side.

\begin{figure*}[t!]
\includegraphics[width=\textwidth]{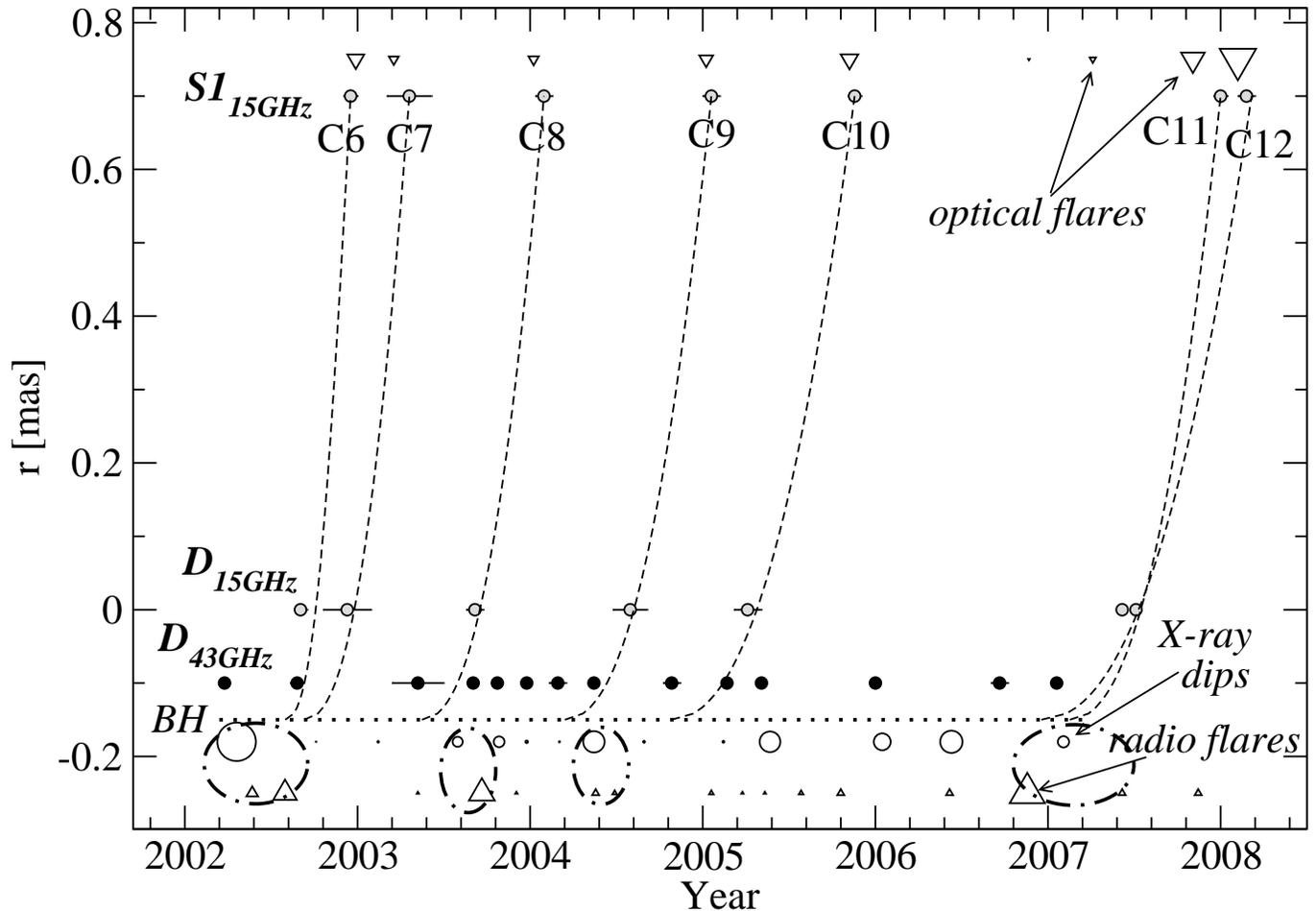}
\caption{Jet kinematics and flaring activity in 3C~120. Black circles
  ($D_{43GHz}$) show the ejection epochs of the jet components
  identified at 43~GHz (Chatterjee et al. 2009). Grey circles show the
  epochs of ejection ($D_{15GHz}$) and passages through S1
  ($S1_{15GHz}$) of the jet components identified at 15~GHz (this
  paper). The dotted line indicates the estimated location of the
  central black hole (BH). Epochs of the X-ray dips (open circles) and
  37~GHz radio flares (upward triangles) from Chatterjee et al. (2009)
  and epochs of the optical flares (downward triangles) are also
  shown. The apparent angular separation between $D_{15GHz}$ and
  $S1_{15GHz}$ is measured from the 15~GHz VLBI images. The locations
  of the core at 43~GHz ($D_{43GHz}$) and the central black hole are
  estimated from the core shift as inferred from the synchrotron
  luminosity of the nuclear jet. Dashed lines show evolution of the
  component separations obtained using the hydrodynamical acceleration
  model of Vlahakis and K\"onigl (2004). Dot-dashed ellipses mark
  possible associations of the 15~GHz components with the X-ray dips
  and 37~GHz flares.}
\label{fig:ej-epochs} 
\end{figure*}

These findings should also be discussed in the context of
relations between the optical and X-ray emission in \objectname{3C~120}
reported recently (Kataoka et al. 2007, Marshall et al. 2009,
Chatterjee et al. 2009, Doroshenko et al. 2009).

Kataoka et al. (2007) argued that the strength of the narrow Fe
K$\alpha$ indicates that a major fraction of the X-ray emission
originates from an unbeamed source, presumably the accretion disk.
This conclusion agrees well with our findings, as our correlations
result from a flaring component in the jet, which carries about 25\%
of the total flux density both in the optical and X-ray regimes. This
is similar to the what is found for \objectname{3C~390.3} (Arshakian
et al. 2009) and \objectname{3C~345} (Schinzel et al. 2010). In the
latter case, the $\Gamma$-ray flux density shows a long-term trend
similar to a trend in the radio flux density of the compact core,
while a (short-term) $\Gamma$-ray flare can be related to a similar
flare in a stationary feature in the radio jet, also coinciding with a
passage of a newly ejected moving jet component. All these results
suggest that the flaring component of the X-ray, optical and radio
emission is different in its nature from the underlying, quiescent
emission associated with the accretion disk, and that it may indeed be
produced by a relativistically moving material in the jet.

Marshall et al. (2009) found a correlation found in \objectname{3C~120}
between the X-ray and optical emission, where X-ray variations were
leading the optical ones by approximately 25 days over a period of
almost three years. Chatterjee et al. (2009) report an even smaller
time delay ($0.5\pm 0.4$\,days) between the X-ray and optical flares.
Doroshenko et al. (2009) have used optical photometry and X-ray light
curves of \objectname{3C~120} over a period of almost 12 years and
found different degrees of correlation between the optical and X-ray
variations present at different periods in the light curves. The sign
of the correlation also depends on the epoch.  All these findings
indicate that the flaring optical and X-ray emission are produced
essentially in the same spatial region. The results of our present
work further support this conclusion and enable identifying the
location of the production site of the flaring component more
accurately, placing it in the relativistic jet, at a $\sim 1.3$\,pc
distance from the central black hole of the AGN.

Dips in the X-ray lightcurve of \objectname{3C~120} are reported to be
followed by appearances of new superluminal components (Marscher et
al. 2002, Chatterjee et al. 2009).  Chatterjee et al. (2009) have also
used 43~GHz VLBI observations and decomposed the 37~GHz light curve of
\objectname{3C~120} into a sequence of exponential flares, in order to
connect the X-ray variations to structural and emission changes in the
radio regime. 

Chatterjee et al. (2009) identified 14 moving jet components, during
the time period of 2002-2007, and do not report a feature that can be
considered a direct counterpart of our stationary feature S1. It
should be noted that the 43~GHz VLBI observations probe a region of
$\sim 1$~mas in extent, which is comparable to the distance between D
and S1 measured at 15~GHz. If the core shift (K\"onigl 1981) between
15 and 43~GHz taken into account (following Lobanov 1998, a shift of
0.1--0.2~mas can be inferred both from an estimate of synchrotron
luminosity of $\sim 2$--$5\times 10^{43}$~erg/s for the compact core,
and from a core shift of 0.9--1.1~mas measured between 2.3 and 8.4~GHz
by Kovalev et al. 2008), S1 is likely to locate close to the outer
edge of the 43~GHz emission detected by Chatterjee et al. (2009). In
this case, its stationarity would be difficult to be recognized from
the 43~GHz data, particularly in presence of subsequent ejections of
moving jet components. More detailed and high-dynamic range
multi-frequency VLBI observations would be needed in order to make a
better assessment of the physical conditions in the flow at distances
of about 1~mas from the 43~GHz core of the jet.

The overall complexity of the light curves and the jet structure may
lead to difficulties of cross-identifying individual events in
different bands. For instance, three radio flares and two jet
identified in Chatterjee et al. (2009) do not have respective
counterparts.  The relative magnitude of different radio flares, X-ray
dips, and brightness of the jet components observed at 43~GHz ({\em
e.g.}, components 04C and 06A) vary by a factor of $\sim 30-60$, with
the weaker events and jet components often being difficult to
cross-identify. It is possible that the weakest flares and X-ray deeps
results from transitory or localised events that do not have a long
lasting effect on the jet and cannot be traced at large
distances. These weaker jet components can, for example, be destroyed
during their passages through the standing shock in S1, and only the
strongest components will propagate through and be detected at large
distances at 15~GHz.

Based on the considerations above, we attempt now to combine our
results with the results from Chatterjee et al. (2009) and to discuss
a scheme that could provide a common framework for the entire
observational data.

Comparison of the peaks of 37~GHz flares (listed in Table~2 in
Chatterjee et al. 2009) and the peaks of the optical flares from our
B-band light curve (see Table~\ref{Table:3C120_times}) yields a time
delay of $0.43\pm 0.14$\,yrs, with the radio flares leading the
optical flares (epochs of the radio and optical peaks are shown in
Figure~\ref{fig:ej-epochs} with upward and downward triangles,
respectively).  We compute the probability that the 37~GHz peaks
shifted by 0.45 yr would be observed to occur randomly within 0.15 yr
of the peaks of the optical flares. The probability that 10 out of
16 shifted peaks of 37~GHz flares coincide by chance with
the optical peaks is $ < 10^{-9}$. It should also be noted that the
37~GHz flares that do not have optical counterparts, are largely
located in the seasonal gaps of the optical light curve.

The time lag between the radio and optical flares (where radio is
leading) is similar to the time taking a moving component of the jet
to travel the distance between D and S1 ($0.45 \pm 0.14$\,yrs).  Such
a succession of the observed flares cannot be easily reconciled with
the overall picture of the shock-in-jet model (Marscher and Gear 1985)
where the flare propagates from higher to lower frequencies as a shock
moves downstream in the jet. One can however attempt to explain such a
behavior by a strong acceleration of the emitting material while it
travels towards S1 (since the peaks of the optical flares are
correlated with the passages of the moving jet components through
S1). We combine together the ejection epochs at 15 and 43~GHz and the
epochs of component passages through S1 (Figure~\ref{fig:ej-epochs})
and plot them together with the times of X-ray dips, and flares in the
radio and optical emission. We apply a 0.1~mas core shift between 15
and 43 GHz and  use this shift to estimate the central black hole
to be located at 0.05 mas further upstream from the core at 43~GHz
(following Lobanov 1998).

Assuming that the jet accelerates hydrodynamically ({\em cf.} Vlahakis
\& K\"onigl 2004) and reaches its terminal speed (characterized by a
Lorentz factor $\Gamma_\mathrm{max}$) at the location of S1, so that
$\Gamma_\mathrm{max}$ can be then inferred for each component from its
apparent proper motion (for the adopted value of $20.5^\circ$ for the
jet viewing angle). We apply the description of Vlahakis \& K\"onigl,
yielding an approximation $\Gamma_\mathrm{j}(r) = 1 +
(\Gamma_\mathrm{max} -1) [r/(r+r_0)]^a$ for the material travelling
along the innermost magnetic field line (assumed to originate from the
accretion disk at a distance $\varpi \approx 10$ Schwarzschild
radii from the central black hole). We find that a characteristic
acceleration distance $r_0 =1.5$~pc (resulting from $\varpi \approx
10^{-4}$~pc) and power index $a=1$ provide a plausible representation for
the evolution of 
component separations in the acceleration zone between the base of the jet
 and S1. The resulting model separations are plotted with dashed
lines in Figure~\ref{fig:ej-epochs}, calculated backwards in time
starting from the epochs of component passages through S1. Note that
the model curves do not go through the ejection points,
$D_\mathrm{15GHz}$, as they represent accelerated
motions, while the ejection epochs are calculated from linear fits
implying a constant speed.

Within this scenario, ejections of all 15~GHz components except C10
can be associated tentatively with strong X-ray dips (circles in
Figure~\ref{fig:ej-epochs}) and strong radio flares at 37~GHz (upward
triangles). It should be further noted that establishing a firm
association of this sort requires a detailed knowledge of the
component trajectories between D and S1, which could be obtained from
a dedicated VLBI monitoring program.  The stronger optical flares
(downward triangles) are well correlated with the component passages
through S1. It should be noted that the terminal speeds for C11 and
C12 are poorly constrained, owing to small number of measurements
available for these components, and it is feasible that their model
ejection epochs may be moved to earlier times. It is not easy to
establish a correspondence with the 43~GHz ejection epochs, but it
should be possible if the trajectories of 43~GHz components are
measured accurately up to a distance of 1~mas from the core and epochs
at which they reach this distance are compared with the epochs of
passages of the 15~GHz components through the region S1. It is
important to verify if the 43~GHz VLBA data reveal any evidence for
accelerated motions at distances $\le 1$~mas, which will constrain the
physical conditions at sub-milliarcsecond scales.

Based on the discussion above, we can suggest that the 15~GHz
components may correspond to the strongest nuclear events, while some
of the 43~GHz components may not survive a passage through a standing
shock in the jet at the location of the stationary feature S1. The
source of the continuum emission is localized not only in the
accretion disk (at the extreme vicinity of the black hole) but also in
the entire acceleration zone of the jet, with strong flares happening
both near the central black hole at D (assumed base of the jet) and at
the standing shock (S1) in the jet. This possibility makes
\objectname{3C~120} one of the prime candidates for studies of flow
acceleration and continuum emission production in radio-loud AGN.
Clearly, a more general and systematic study of relation between the
radio, optical and X-ray emission in \objectname{3C~120} is strongly
justified.

The correlation between the optical flares and the passages of moving
jet components through the location of the stationary feature S1 is
best interpreted in terms of disturbances (plasma condensations)
propagating in the relativistic flow (Arshakian~et~al.~2009). In this
scenario, the optical continuum generated in \objectname{3C~120} is
dominated by non-thermal, Doppler-boosted synchrotron emission from a
relativistically accelerated plasma in the jet. The jet can be
collimated and accelerated on scales of $10^5$--$10^6\,R_\mathrm{g}$
(where $R_\mathrm{g}$ is the gravitational radius of the central black
hole) by an MHD mechanism (e.g. Vlahakis \& K\"onigl 2004), with
emission peaking at the optical wavelength during a passage through a
standing shock compressing the moving material and enhancing
substantially the magnetic field. As a moving plasma condensation
separates from the standing shock, it appears as a moving knot in the
flow. From stellar velocity dispersion measurements of 3C~120 optical
spectra, the BH mass is estimated to be $\sim 3.5 \times 10^{7}$
M$_{\sun}$ (Greene \& Ho 2006). Thus, the distance of component S1
from the core of the jet corresponds to $\sim 9 \times 10^{5}$
$R_\mathrm{g}$. This scale is consistent with the end of the
acceleration and collimation zone in relativistic jets.  Further
studies, in close detail, of the evolution of compact radio emission
during radio and optical flares in objects like \objectname{3C~120} is
essential for understanding the mechanism for acceleration and energy
release in powerful relativistic jets and its relation to generation
of non-thermal optical continuum and broad line emission in radio-loud
AGN.

\section{CONCLUSIONS}
To investigate the link between optical continuum variability and
subparsec-scale jet in the radio-loud galaxy 3C\,120, we combined the
radio VLBI (15\,GHz) and optical photometry (B-band) data observed
during the period of nearly eight years. We found a significant
correlation between the jet kinematics on parsec-scales and optical
flares on scales from several months to about two years. All optical
flares are associated with the jet ejection events: the flare rises
after the epoch of ejection of a new jet component (at D) and it
reaches the maximum around the epoch at which the ejected radio knot
passes the stationary radio component (S1) downstream the jet. The
radio passages of new components through S1 delay the maxima of
optical flares on the average by $\la 0.1$ yr. These results confirm
correlations found in Arshakian~et~al.~(2008,~2009) for the radio
galaxy 3C\,390.3 and support the idea that the correlation between
optical flares and kinematics of the jet could be a common feature for
all radio-loud galaxies and quasars.

The link between optical continuum variability and kinematics of the
parsec-scale jet is interpreted in terms of optical flares generated
by disturbances in the jet flow while moving from the stationary
component D of the jet to the stationary component S1 located at a
distance of about one parsec. Modeling of optical flares as
synchrotron flares caused by density variations in the relativistic
flow showed that all optical flares require at most a factor of seven
increase of the particle density. The variation of the particle
density at subparsec-scales should be smaller if consider a more
realistic model in which the magnetic field and/or Lorentz factor of
the jet increase downstream of the jet, in the
acceleration/collimation zone, between D and S1. In this scenario, the
correlated X-ray and optical continuum emission are produced in the
relativistic jet at some distance ($< 1$ pc) from the central black
hole rather than in the accretion disk. \\

Establishing such a relation may have strong implications for the
physics of the central regions in radio-loud AGN and AGN in general
(see also Arshakian et al. 2009). The link between the optical
continuum and radio jet challenges the existing models in which the
optical continuum and broad-line emission are both localized around
the disk or near the central black hole of an AGN. It also questions
the common assumptions about the central engine used, in particular,
for the reverberation mapping technique (Peterson et al. 2002), which
combines the broad emission line width with the size of the
line-emitting region estimated from the optical continuum
luminosity. In 3C\,120, a substantial fraction of the ionizing
continuum is produced in the relativistic jet and it illuminates
thermal material most likely organized in a sub-relativistic
outflow. The time delays and profile widths measured in radio-loud AGN
may reflect not only the Keplerian motion of the emitting line gas,
but also an outflowing component of the gas accelerated by
non-gravitational forces. This can lead to large errors in estimates
of black hole masses made from monitoring of the broad emission lines.

In order to address these issues, we are carrying out a long-term
spectropolarimetry campaign for the \objectname{3C~120} and several
other nearby radio galaxies aimed to determine the amount of
non-thermal optical continuum and to tackle evidence for non-virial
motions in the BLR. Follow-up high resolution radio observations at
other frequencies and epochs, can help to characterize the nature of
the components D and S1.

We thank anonymous referees for a number of useful comments and
suggestions which significantly improved this manuscript.  This work
was supported by CONACYT research grant 54480 (Mexico) and the Russian
Foundation for Basic Research (Grant No. 09-02-01136a) .  JLT
acknowledges support from the CONACYT program for PhD studies, the
International Max-Planck Research School for Radio and Infrared
Astronomy at the Universities of Bonn and Cologne and the Deutscher
Akademischer Austausch Dienst (DAAD) for a short-term scholarship in
Germany. This research has made use of data from the MOJAVE database
that is maintained by the MOJAVE team (Lister et al. 2009a). The VLBA
is an instrument of the National Radio Astronomy Observatory, a
facility of the National Science Foundation operated under cooperative
agreement by Associated Universities, Inc.


\begin{thebibliography}{}

\bibitem[Alberdi et al. (2000)]{alberdi02} Alberdi, A., G\'omez, J.~L.,
Marcaide, J.~M., Marscher, A.~P., P\'erez-Torres, M.~A. 2000, \aap, 361,
529

\bibitem[Arshakian et al.(2010)]{arshakian10} Arshakian, T.~G.,
Le{\'o}n-Tavares, J., Lobanov, A.~P., Chavushyan, V.~H., Shapovalova, A.~I.,
Burenkov, A.~N., \& Zensus, J.~A.\ 2010, \mnras, 401, 1231



\bibitem[Arshakian et al.(2008)]{arshakian08} Arshakian, T.~G.,
Lobanov, A.~P., Chavushyan, V.~H., Shapovalova, A.~I.,
\& Zensus, J.~A.\ 2008, Relativistic Astrophysics Legacy and Cosmology - Einstein's, 189,
arXiv:astro-ph/0602016.

\bibitem[Belokon (1987)]{belokon87}
Belokon, E. T. 1987, Astrophysics, 27, 588

\bibitem[Becker et al. (2005)]{becker05} Becker, P. A., Das, S., Le,
T. 2008, \apj, 677, L93

\bibitem[Chatterjee et al. (2009)]{chatt09} 
Chatterjee,  R., Marscher, A. P., Jorstad,  S. G., Olmstead,  A. R., McHardy,  I. M. et al. 2009, accepted to \apj, arXiv:0909.2051


\bibitem[Cornwell \& Fomalont (1999)] {cornwell1999} Cornwell, T.,
Fomalont, E.B. 1999, in ``Synthesis Imaging in Radio Astronomy II'',
eds. G. B. Taylor, C. L. Carilli, and R. A. Perley. ASP Conference
Series, Vol. 180, (ASP: San Francisco), p. 187.

\bibitem[Daly et al.(1988)]{daly88} 
Daly, R. A., Marscher, A. P. 1988, \apj, 334, 539

\bibitem[D'arcangelo et al.(2007)]{darcangelo07} 
  D'Arcangelo, F. D., Marscher, A. P., Jorstad, S. G., Smith, P. S.,
  Larionov, V. et al.\ 2007, \apj, 659, L10

\bibitem[Doroshenko et al. (2005)]{soroshenko05}
  Doroshenko, V. T., Sergeev, S. G., Merkulova, N. I., Sergeeva,
  E. A., Golubinsky et al. 2005, Ap, 48, 304

\bibitem[Doroshenko et al. (2009)]{doroshenko09}
Doroshenko, V. T., Sergeev, S. G, Efimov, Y. S, Klimanov, S. A. and Nazarov S. V. 2009,AstL, 35, 361

\bibitem[Fomalont (1999)]{fomalont99}
  Fomalont, E.~B., 1999, in ASPConf.Ser., v.180, synthesis imaging in
  radio astronomy II, eds. Taylor, G.B., Carilli, C.L., Perley,
  R.A. (ASP: San Francisco),301

\bibitem[G\'omez et al.(1995)]{gomez95} 
  G\'omez, J. L., Marti, J. M. A., Marscher, A. P., Ibanez, J. M. A.,
  Marcaide, J. M.,\ 1995, ApJL, 449, L19

\bibitem[G\'omez et al.(1997)]{gomez97} 
  G\'omez, J. L., Marti, J. M. A., Marscher, A. P., Ibanez, J. M. A.,
  Alberdi, A.,\ 1997, ApJ, 482, L33

\bibitem[G\'omez et al.(2001)]{gomez01} 
  G\'omez, J. L., Marscher, A. P., Alberdi, A., Jorstad, S. G., Agudo,
  I.,\ 2001, ApJ, 561, 161


\bibitem[Greene \& Ho (2006)]{gho06}
Greene, J. E., \& Ho, L. C. 2006, \apj, 641, L21
 
\bibitem[Jorstad et al. (2005)]{jorstad05}
  Jorstad, S. G., Marscher, A. P., Lister, M. L., Stirling, A. M.,
  Cawthorne, T. V. et al., 2005, \aj, 130, 1418

\bibitem[Kellermann et~al. (1998)]{keller98}
{Kellermann}, K.~I., {Vermeulen}, R.~C., {Zensus}, J.~A., \& {Cohen}, M.~H.
  1998, \aj, 115, 1295

\bibitem[Kellermann et~al. (2004)]{keller04}
  Kellermann, K. I., Lister, M. L., Homan, D. C., Vermeulen, R. C.,
  Cohen, M. H et al.  2004, \apj, 609, 539

\bibitem[K\"onigl (1981)]{konigl81}
   K\"onigl, A. 1981, \apj, 243, 700

\bibitem[Kovalev et al. (2008)]{kovalev2008} Kovalev, Y. Y., Lobanov, A. P.,
Pushkarev, A. B., Zensus, J. A. 2008, A\&A, 483, 759

\bibitem[{Lister} (2003)]{lister03}
{Lister}, M.~L. 2003, in ASP Conf. Ser. 300, Radio AStronomy at the Fringe,
ed. J.A. Zensus, M.H. Cohen, \& E. Ros (San Francisco:ASP),71

\bibitem[{{Lister} \& {Homan}(2005)}]{lister05}
{Lister}, M.~L., \& {Homan}, D.~C. 2005, \aj, 130, 1389


\bibitem[Lister et al. (2009)]{lister09a}
  Lister, M. L., Aller, H. D., Aller, M. F. et al. 2009, \aj, 137, 3718

\bibitem[Lister et al. (2009b)]{lister09b} 
Lister, M. L., Cohen, M. H., Homan, D. C. et al. 2009, \aj, 138, 1874

\bibitem[Lobanov (1996)]{lobanov96}
 Lobanov, A.~P., 1996, PhD Thesis, NMIMT, Socorro, USA

\bibitem[Lobanov(1998)]{lobanov98} Lobanov A.~P.,\ 1998, A\&A, 330, 79

\bibitem[{{Lobanov}(2005)}]{andrei05} {Lobanov}, A.~P., 2005,
astro-ph/0503225

\bibitem[Lobanov \& Zensus (1999)]{lobanov99} Lobanov, A.P., Zensus,
J.A. 1999, ApJ, 521, 509

\bibitem[Mandal \& Chakrabarti (2008)]{mandal08} Mandal, S.,
Chakrabarti, S.~K. 2008, \apj, 689, L17

\bibitem[Marshall et al. (2009)]{marshall09}
  Marshall, K., Ryle, W. T., Miller, H. R., Marscher, A. P., Jorstad,
  S. G. et al. 2009, \apj, 696, 601

\bibitem[Marscher\& Gear (1985)]{marscher85}
Marscher A. P. \& Gear, W. K. 1985, ApJ, 298, 114

\bibitem[Marscher et al.(2002)]{marscher02}
  Marscher, A. P., Jorstad, S. G., G\'omez, J. L., Aller, M. F.;
  Ter\"asranta, H., et al. \ 2002, Nature, 417, 625


\bibitem[Marscher et al.(2008)]{marscher08}
  Marscher, A. P.; Jorstad, S. G., D'Arcangelo, F. D., Smith, P. S.,
  Williams, G. G., et al. \ 2008, Nature, 452, 966

\bibitem[Pearson(1997)]{pearson} Pearson, T. J.,\ 1997, in ASP
  Conf. Ser. 180, Synthesis imaging in radio astronomy II, eds.
  G.~B.~Taylor, C.~R.~Carilli, R.~A.~Perley, 335

\bibitem[Perucho \& Mart\'i (2007)]{perucho07} Perucho, M., Mart\'i,
J.~M. 2007, MNRAS, 382, 526

\bibitem[Peterson et al.(2002)]{peterson02} Peterson, B. M., 2002,
  Advanced Lectures on The Starburst-AGN Connection, eds Aretxaga I.,
  Kunth D. \& M\'ujica R., Singapore World Scientific, 3

\bibitem[Romanova \& Lovelace(1996)]{romanova96} 
  Romanova, M. M., Lovelace, R. V. E.,\ 1996, A\&AS, 120, 583

\bibitem[Savolainen et al. (2006)]{savolainen02}
 Savolainen et al. 2006, \aap, 446, 71

\bibitem[Sergeev et~al. (2005)]{sergeev05}
  Sergeev, S. G., Doroshenko, V. T., Golubinskiy, Y. V., Merkulova,
  N. I., \& Sergeeva, E. A. 2005, \apj, 622, 129

\bibitem[Shapovalova et~al. (2001)]{shap01}
  Shapovalova, A. I., Burenkov, A. N., Carrasco, L., Chavushyan,
  V. H., Doroshenko, V. T., et~al. 2001, \aap, 376, 775

\bibitem[Schinzel et al. (2010)]{schinzel10} Schinzel, F., Lobanov, A.~P., Zensus, J.~A. {\em (in prep.)}

\bibitem[Soldi (2008)]{soldi08}
  Soldi, S., T\"urler, M., Paltani, S., Aller, H. D., Aller, M. F. et
  al. 2008, \aap, 486, 41

\bibitem[Tyul'bashev \& Chernikov (2004)]{tyulbashev04} Tyul'bashev, S. A., 
Chernikov, P. A. 2004, Astronomy Reports, 48, 716

\bibitem[Valtaoja et al. (1999)]{valtaoja99}
Valtaoja, E. et al. 1999, \apjs, 120, 95

\bibitem[Vlahakis \& K\"onigl (2004)]{vlahakis2004} 
  Vlahakis, N. \&  K\"onigl, A.  2004, \apj, 605, 656

\end{thebibliography}
\end{document}